\documentclass[aps,graphicx,6pt,showkeys,showpacs,twocolumn]{revtex4}
\usepackage{amsmath}
\usepackage{amscd}
\usepackage{graphicx}
\usepackage{subfigure}
\usepackage{appendix}

\begin{document}
\title[Short Title]{Reverse
engineering of a non-lossy adiabatic Hamiltonian for non-Hermitian
systems}

\author{Qi-Cheng Wu$^{1}$}
\author{Ye-Hong Chen$^{1}$}
\author{Bi-Hua Huang$^{1}$}
\author{Yan Xia$^{1,}$\footnote{E-mail: xia-208@163.com}}
\author{Jie Song$^{2}$}
\affiliation{$^{1}$Department of Physics, Fuzhou University, Fuzhou 350002, China\\
             $^{2}$Department of Physics, Harbin Institute of Technology, Harbin 150001, China}

\begin{abstract}
We generalize the quantum adiabatic theorem to the non-Hermitian
system and build a strict adiabaticity condition to make the
adiabatic evolution non lossy when taking into account the effect
of adiabatic phase. According to the strict adiabaticity
condition, the non-adiabatic couplings and the effect of the
imaginary part of adiabatic phase should be eliminated as much as
possible. Also the non-Hermitian Hamiltonian reverse engineering
method is proposed for  adiabatically driving an artificial
quantum state. Concrete two-level system is adopted  to show the
usefulness of the reverse engineering method.  We obtain the
desired target state by adjusting extra rotating magnetic fields
at a predefined time. Furthermore, the numerical simulation shows
that certain noise and dissipation in the systems are no longer
undesirable, but play a positive role in the scheme. Therefore,
the scheme is quite useful for quantum information processing in
some dissipative systems.
\end{abstract}


\maketitle

\footnotesize

\section{Introduction}
As an essential task in many areas of quantum information science
ranging from quantum information
processing~\cite{Barz2013,Yuprl2004} and coherent manipulation of
quantum systems~\cite{Scully1997}  to high-precision
measurements~\cite{Kasevich2002,Kotru2015}, the quantum-state
engineering
(QSE)~\cite{jgMuga,Duzzioni2011,wsl2015,Jingpra2013,Yupra1999,Chenprl2010,Liang2015},
has attracted much attention, which promotes the development of
experimental technique and theoretical scheme. The quantum
adiabatic theorem (QAT), an important way of realizing QSE, has
been widely studied and the basic properties of the QAT are being
scrutinized both theoretically and
experimentally~\cite{Berry1984,Simon1983,Farhi2001,Campo2011,Unanyan1999,Simon2011,Bohm2003}.
The basic idea of  QAT can be summarized as follows: if the
control parameters in the time-dependent Hamiltonian change
slowly, the system will follow closely along an eigenstate
trajectory up to a adiabatic phase factor when it is initially in
one of the eigenstates. Thereinto, the adiabatic phase is a
complicated factor which can be divide into dynamical phase and
geometrical phase~\cite{Berry1984}. Interestingly, in the
Hermitian adiabatic Hamiltonians scenario, one can focus on the
dynamics of the eigenstate, and neglect the complicated phase
factor since it can be discarded as the common pure phase.
However, in practice the quantum system  inevitably interacts with
the surrounding environment, e.g. the non-Hermitian (NH)
systems~\cite{Garrison1988,Dattoli1990,Bender,Graefe2008,Miniatura1990,Sun1993,Dridi2010,Dridi2012,Mugapra2014,Berry2011,Lavdjpa2012,Chenpra2012,Torosov2013}.
In this case, the complicated adiabatic phase factor could not be
simply discarded as the common pure phase any more, since it
generally is not a pure (real) phase factor. Then, the ideal
robustness and the intended dynamics may be spoiled by the
accumulation of the imaginary part of the adiabatic phase due to
noise and undesired interactions. Thus,  it is very worthwhile to
look for the novel methods which are robustness, and figure out
the strict adiabaticity condition for NH Hamiltonians when taking
into account the effect of adiabatic phase.

In fact, several authors have  paid attention to the study of
adiabaticity in NH
systems~\cite{Miniatura1990,Sun1993,Dridi2010,Dridi2012,Mugapra2014}.
For example, Miniatura \textit{et al.} have set a rough estimate
of an adiabaticity condition by analogy with the Hermitian
counterpart and  recognized the importance of  the nonadiabatic
transition~\cite{Miniatura1990}. Subsequently, Sun has devoted to
the generalization of the high-order adiabatic approximation
method for the NH quantum systems by using perturbation theory and
integration by parts, and obtained an adiabaticity condition
similar to the Hermitian one with the damping factor and the
oscillating factor~\cite{Sun1993}. Recently, Dridi \textit{et al.}
have established a generalization of the Davis-Dykhne-Pechukas
formula by the complex time method and showed a general adiabatic
approximation for lossy two-state
models~\cite{Dridi2010,Dridi2012}. More recently, Ib\'{a}\~{n}ez
and Muga have generalized  the concept of population for NH
systems to characterize adiabaticity and worked out an approximate
adiabaticity criterion~\cite{ Mugapra2014}. Indeed, the
adiabaticity of a given NH system has been discussed well  with
those excellent
methods~\cite{Miniatura1990,Sun1993,Dridi2010,Dridi2012,Mugapra2014}.
However, in principle above methods didn't give a clear
quantitative analysis about the dynamics of the bare state in the
eigenstate and the adiabatic phase. In some cases, we may observe
a false adiabaticity due to the problematic or obscure population
concept~\cite{Berry2011,Lavdjpa2012,Chenpra2012}. In addition,
they will also be limited severely by the presence of the strong
dissipation effects in some applications~\cite{Lavdjpa2012,
Mugapra2014}. Above problems make designing perfect scheme to
reach the intended dynamics for the NH systems very challenging.

In this paper, we will introduce a novel method to solve the
problems shown above. Different from the previous
schemes~\cite{Sun1993,Dridi2012,Mugapra2014} which are proposed to
explore the adiabatic approximation condition for a given NH
system, we are dedicated to setting a strict adiabaticity
condition to make the adiabatic evolution non lossy when taking
into account the effect of adiabatic phase, and exploring the
Hamiltonian which will exactly satisfy the strict condition via
the reverse engineering method. The scheme has following
advantages: (1) We take the effect of adiabatic phase into
consideration, and make the adiabatic process be a strict one
without lossy in NH system. (2) By using the reverse engineering
method, we can design the Hamiltonian  to realize the intended
dynamics according to the demand. (3) The noise or certain
dissipation in the systems can do positive job in the scheme. We
can obtain the desired target state by adjusting extra rotating
magnetic fields at a predefined time even in the dissipative
system. Therefore, the scheme makes it possible to realize the QSE
for some dissipative systems.

The rest of this paper is arranged as follows. In
Sec.~\ref{section:II}, we briefly review some important properties
of the NH Hamiltonians, and build a strict adiabaticity condition
which contains two parts: an auxiliary adiabaticity condition with
respect to adiabatic phase and a general adiabatic condition given
via the Feshbach P-Q partitioning
technique~\cite{LAWuprl2009,Jingpra2014,Gaspard1999}. In
Sec.~\ref{section:III}, we explicitly discuss how to engineer the
NH Hamiltonian which could exactly satisfy the strict adiabaticity
condition. Then, we consider a concrete two-level system example
to show the usefulness of our reverse engineering method. Both
experimental feasibility and population engineering are discussed
step by step in Sec.~\ref{section:IV}. Finally we give a summary
 in Sec.~\ref{section:V}.

\section{Basic theories}\label{section:II}
\subsection{NH Hamiltonians: basic formulas }

For NH systems, the usual approximations and criteria  are not
necessarily valid, so the results which are applicable for
Hermitian systems have to be reconsidered and modified. We first
briefly recall some important properties of the NH
Hamiltonians~\cite{Garrison1988,Dattoli1990}. Consider an
arbitrary  time-dependent NH Hamiltonian $H(t)$ with $N$
nondegenerate instantaneous eigenstates
$\{|{\phi_{n}}(t)\rangle\}$, $n$=$1,2,...,N.$ It satisfies
following eigenvalue equation
\begin{eqnarray}\label{eq0-1}
  H(t)|{\phi_{n}}(t)\rangle=E_{n}(t)|{\phi_{n}}(t)\rangle.
\end{eqnarray}
As the adjoint operator of $H(t)$, $H(t)^\dag$, will also satisfy
following eigenvalue equation
\begin{eqnarray}\label{eq0-2}
  H(t)^\dag|\widehat{{\phi_{n}}}(t)\rangle=E_{n}^{*}(t)|\widehat{{\phi_{n}}}(t)\rangle,
\end{eqnarray}
where $\{|\widehat{{\phi_{n}}}(t)\rangle\}$ are the instantaneous
eigenstates of $H(t)^\dag$ and also the biorthogonal partners of
$\{|{\phi_{n}}(t)\rangle\}$,  the asterisk means complex
conjugate. The biorthogonal partners are normalized to satisfy the
biorthogonality relation
\begin{eqnarray}\label{eq0-3}
\langle\widehat{{\phi_{n}}}(t)|\phi_{m}(t)\rangle=\delta_{nm},
\end{eqnarray}
and the closure relation
\begin{eqnarray}\label{eq0-4}
\sum_{n}|\widehat{{\phi_{n}}}(t)\rangle\langle\phi_{n}|=\sum_{n}|{\phi_{n}}(t)\rangle\langle\widehat{{\phi_{n}}}(t)|=1.
\end{eqnarray}
With above properties, the Hamiltonian and its adjoint can be
rewritten as
\begin{eqnarray}\label{eq0-5}
H(t)&=&\sum_{n}|{\phi_{n}}(t)\rangle{E_{n}(t)}\langle\widehat{{\phi_{n}}}(t)|,
\cr
H(t)^\dag&=&\sum_{n}|\widehat{{\phi_{n}}}(t)\rangle{E_{n}^{*}(t)}\langle{\phi_{n}}(t)|.
\end{eqnarray}

\subsection{The auxiliary
adiabaticity condition for the NH systems with respect to
adiabatic phase}

According to the adiabatic theorem, a state with initial condition
$|\phi(0)\rangle$=$|{\phi_{n}}(0)\rangle$  will evolve
adiabatically if its dynamics is well approximated by
$|\phi(t)\rangle$=$e^{i\beta_{{n}}(t)}|{\phi_{n}}(t)\rangle$.
Furthermore, if $|{\phi_{n}}(t)\rangle$ is the instantaneous state
of the system Hamiltonian $H(t)$  and $|\phi(t)\rangle$ satisfies
the Schr\"{o}dinger equation ($\hbar$=$1$)
\begin{eqnarray}\label{eq0-8}
i|\dot{\phi}(t)\rangle=H(t)|\phi(t)\rangle,
\end{eqnarray}
we can obtain the adiabatic phase
\begin{eqnarray}\label{eq0-9}
\beta_{{n}}(t)&=&\int_{0}^{t}[-E_{n}(t')+i\langle\widehat{{\phi_{n}}}(t')|{\dot{\phi}}_{n}(t{'})\rangle]dt{'}.
\end{eqnarray}
However, this ansatz of the adiabaticity for the NH system is not
strict. The imaginary part of the adiabatic phase will induce the
decay of system  and cause confusion about the validity of the
adiabaticity. Consequently, it's necessary to forcibly eliminate
Im$[\beta_{{n}}(t)]$ to keep the adiabatic scheme working well,
that is, we should insure
\begin{eqnarray}\label{eq0-10}
-\textrm{Im}[E_{n}(t)]+\textrm{Re}[\langle\widehat{{\phi_{n}}}(t)|{\dot{\phi}}_{n}(t)\rangle]=0.
\end{eqnarray}
Then, the adiabatic phase can be safely discarded as a common pure
phase when we investigate the dynamics of the target state, even
in the NH Systems. Notice that, Eq.~(\ref{eq0-10}) is the
auxiliary adiabaticity condition which allows one to make the
adiabatic evolution non lossy with respect to adiabatic phase,
which is the primary result to be used in following work.

\subsection{ The general adiabatic condition for the NH systems}\label{section:my}

In general, a state at time $t$ can be expressed as
\begin{eqnarray}\label{eq0-13}
|\Psi(t)\rangle&=&\sum_{n}{\Psi_{n}(t)}e^{i\beta_{{n}}(t)}|{\phi_{n}}(t)\rangle,
\end{eqnarray}
where the phase factor $\beta_{{n}}(t)$ satisfies
Eq.~(\ref{eq0-9}) for arbitrary $n$ and $\Psi_{n}(t)$ is
considered as a complex function. It's obvious that $\Psi_{n}(t)$
is the key coefficient associated with the dynamics of
$|{\phi_{n}}(t)\rangle$. Therefore, an exact dynamical equation
for $\Psi_{n}(t)$ is highly desirable. Assuming $|\Psi(t)\rangle$
satisfies the Schr\"{o}dinger equation, we can obtain the
following  equations,
\begin{eqnarray}\label{eq0-14a}
i\dot{\Psi}_{n}(t)&=&-i\sum_{m\neq{n}}{\langle\widehat{{\phi_{n}}}(t)|\dot{\phi}_{m}(t)\rangle}e^{i(\beta_{{m}}(t)-\beta_{{n}}(t))}\Psi_{m}(t)
\cr&=&\sum_{m\neq{n}}H'_{mn}\frac{\Psi_{m}(t)}{\Psi_{n}(t)}\Psi_{n}(t),
\end{eqnarray}
\begin{eqnarray}\label{eq0-14b}
i|\dot{\Psi}'_{n}(t)\rangle&=&\sum\limits_{m\neq{n}}H'_{mn}|\Psi_{m}(t)\rangle\langle{\Psi_{n}(t)}|\Psi'_{n}(t)\rangle
\cr&=&H{'}(t)|\Psi'_{n}(t)\rangle,
\end{eqnarray}
where
$H'_{mn}$$\equiv$$-i\langle\widehat{{\phi_{n}}}(t)|\dot{\phi}_{m}(t)\rangle
e^{i(\beta_{m}(t)-\beta_{n}(t))}$. According to adiabatic theorem,
$|{\phi_{n}}(t)\rangle$ will evolve adiabatically if the term on
the left-hand side of Eq.~(\ref{eq0-14a}) approaches to zero.
Moreover, it is interesting to find that the form of
Eq.~(\ref{eq0-14a}) is similar to the form of the artificial
Schr\"{o}dinger equation Eq.~(\ref{eq0-14b}) for the vector
$|\Psi'_{n}(t)\rangle$=$[\Psi_{1}(t),\Psi_{2}(t),\Psi_{3}(t),...\Psi_{n}(t)]^{T}$
(the superscript $T$ denotes the transpose operator) with the
rotating representation Hamiltonian $H'(t)$.  So we can deal with
$\Psi_{n}(t)$ with the help of Eq.~(\ref{eq0-14b}). In fact,
$H'(t)$ describes the coupling transitions between the
instantaneous eigenstates {$\{|\phi_{n}(t)\rangle\}$}, the
so-called non-adiabatic couplings.

We should stress that in this paper we don't intend to research
fully adiabatic dynamics (for all modes). The problem we address
is that the adiabatic dynamics of one target component (for one
mode). Without loss of generality, the target component can be
denoted as $\Psi_{1}(t)$, corresponding to the target eigenstate
$|\phi_{1}(t)\rangle$ of $H(t)$. In order to obtain a better
understanding of the adiabatic dynamics of $\Psi_{1}(t)$, the
Feshbach P-Q partitioning technique \cite{LAWuprl2009,Jingpra2014}
is introduced. According to the P-Q partitioning technique, the
state $|\Psi'_{n}(t)\rangle$ and the rotating representation
Hamiltonian $H'(t)$ in the Schodinger equation Eq.~(\ref{eq0-14b})
can be always partitioned into the following form,
\begin{eqnarray}\label{eq0-15}
|\Psi'_{n}(t)\rangle&=& \Big[\frac{P}{Q} \Big], \ \ H'(t)=
\left(\begin{array}{ccc}
 0 & R \\
 W & D \\
\end{array}\right),
\end{eqnarray}
where $P$ associated with the target state is equal to
$\Psi_{1}(t)$, while Q associated with $(N$-$1)$-dimensional
vector denotes the rest of the state spaces. The vector
$R$$\equiv$$[R_{2},R_{3}...R_{n}]$ with
$R_{m}$=$-i\langle\widehat{{\phi_{1}}}(t)|\dot{\phi}_{m}(t)\rangle
e^{i(\beta_{m}(t)-\beta_{1}(t))}$ $(m$$\geq$$2)$, while, the
vector $W$$\equiv$$[W_{2},W_{3}...W_{n}]^T$ with
$W_{m}$=$-i\langle\widehat{{\phi_{m}}}(t)|\dot{\phi}_{1}(t)\rangle
e^{i(\beta_{1}(t)-\beta_{m}(t))}$. The
$(N$-$1)$$\times$$(N$-$1)$-matrix
$D$$\equiv$$\sum\limits_{m\neq{n}}D_{mn}|\Psi_{m}(t)\rangle\langle{\Psi_{n}(t)}|$,
where
$D_{mn}$=$-i\langle\widehat{{\phi_{n}}}(t)|\dot{\phi}_{m}(t)\rangle
e^{i(\beta_{m}(t)-\beta_{n}(t))} (m,n$$\geq$$2)$.

Substituting Eq.~(\ref{eq0-15}) into Eq.~(\ref{eq0-14b}), we
obtain the following equations
\begin{eqnarray}\label{eq0-16}
i\dot{P}=RQ,  \      \     \  i\dot{Q}=WP+DQ.
\end{eqnarray}
The formal solution of Eq.~(\ref{eq0-16}) can be written as
\begin{eqnarray}\label{eq0-17}
i\dot{P}=-iR(t)\int_{0}^{t}G(t,s)W(s)P(s) ds+R(t)G(t,0)Q(0),
\end{eqnarray}
where
$G(t,s)$=$\mathcal{T}_{\leftarrow}\{\exp[-i\int_{s}^{t}D(s')ds']\}$
is the time-ordered evolution operator. Under the condition
$P(0)$=$1$ and $Q(0)$=$0$, we have the exact dynamical equation
for the $P$ part
\begin{eqnarray}\label{eq0-18}
\dot{P}=-R(t)\int_{0}^{t}G(t,s)W(s)P(s) ds=-\int_{0}^{t}g(t,s)P(s)
ds,
\end{eqnarray}
where $g(t,s)$=$R(t)G(t,s)W(s)$ is an effective propagator which
plays a very important role in the analysis of adiabaticity.
Notice that, the general adiabatic approximation condition is
$\int_{0}^{t}g(t,s)P(s) ds$=$0$, that is, the propagator
$g(t,s)$=$0$ or $g(t,s)$ is factored by a rapid oscillating
function \cite{Jingpra2014,Marzlin2004}, which is also the primary
result to be used in following work.

For an effective two-level system, the associated rotating
representation Hamiltonian $H{'}(t)$ reads
\begin{eqnarray}\label{eq0-19}
H'(t)=-i\left(\begin{array}{ccc}
 0 & \langle\widehat{{\phi_{1}}}(t)|\dot{\phi}_{2}(t)\rangle e^{i\Delta \beta(t)} \\
\langle\widehat{{\phi_{2}}}(t)|\dot{\phi}_{1}(t)\rangle e^{-i\Delta \beta(t)} & 0 \\
\end{array}\right),
\end{eqnarray}
where $\Delta \beta(t)$=$\beta_{2}(t)-\beta_{1}(t)$. When the
effective two-level system is initially in the eigenstate
$|{\phi_{1}}(0)\rangle$, the propagator $g(t,s)$ reads
\begin{eqnarray}\label{eq0-20}
g(t,s)=-\langle\widehat{{\phi_{1}}}(t)|\dot{\phi}_{2}(t)\rangle
\langle\widehat{{\phi_{2}}}(s)|\dot{\phi}_{1}(s)\rangle
e^{i\int_{s}^{t}(\dot{\beta}_{2}(s{'})-\dot{\beta}_{1}(s{'}))ds{'}}.
\end{eqnarray}
Notice that, Eqs. (\ref{eq0-19}) and (\ref{eq0-20}) are also the
primary results to be used in following work.

\section{THE NH HAMILTONIAN  REVERSE  ENGINEERING METHOD AND APPLICATIONS}\label{section:III}
\subsection{The NH Hamiltonian reverse engineering method}

In this section, we will start with an engineering method about
how to engineer the yet unknown NH Hamiltonian which could exactly
satisfy the strict adiabaticity condition. From the special
properties of the NH Hamiltonian [see Eq.~(\ref{eq0-5})], one can
conclude that the design process can be divided into two steps:
designing the eigenvectors and modifying the eigenvalues. Here we
should make some remarks on the eigenvectors designs. (1) The goal
of our scheme is driving the eigenvectors of an initial
Hamiltonian into those of a final Hamiltonian, so the designed
eigenvectors must connect the initial state with the target state.
(2) Our scheme is working in the NH Hamiltonians scenario, the
eigenvectors must satisfy the biorthogonality relation and the
closure relation. (3) The eigenvectors must evolve adiabatically,
that is, they should satisfy the general adiabatic condition which
has been discussed in Sec.~\ref{section:my}. Once the eigenvectors
designs are completed, we can reconsider and modify the
eigenvalues resorting to normalization ambiguities in the
eigenvectors of NH Hamiltonians. More specially, we should
consider the auxiliary adiabaticity condition with respect to
adiabatic phase for the new eigenvector in this step.

Before the elaborating on manipulating a two-level system to the
target state, we will give a simple restriction on eigenvectors to
satisfy the biorthogonality relation and the closure relation from
the view of mathematics. Without loss of generality, for a $n$
dimensions system, we assume the eigenstates
$\{|{\phi_{n}}(t)\rangle\}$ of $H(t)$ read
\begin{eqnarray}\label{eq2-1}
|\phi_{1}(t)\rangle&=&A_{11}(t)|1\rangle+A_{21}(t)|2\rangle+\cdots+A_{n1}(t)|n\rangle,\cr
|\phi_{2}(t)\rangle&=&A_{12}(t)|1\rangle+A_{22}(t)|2\rangle+\cdots+A_{n2}(t)|n\rangle,\cr
&&\vdots\cr
|\phi_{n}(t)\rangle&=&A_{1n}(t)|1\rangle+A_{2n}(t)|2\rangle+\cdots+A_{nn}(t)|n\rangle,
\end{eqnarray}
where $|l\rangle$ ($l$=$1,2,3\cdots n$) is the bare state for the
system and $A_{jk}(t)$ ($j,k$=$1,2,3\cdots n$) is a devisable
function associated with the bare state $|j\rangle$ in
$|\phi_{k}(t)\rangle$. In a similar manner,  the biorthogonal
states of $\{|{\phi_{n}}(t)\rangle\}$  are expressed as
\begin{eqnarray}\label{eq2-2}
\langle\widehat{{\phi_{1}}}(t)|&=&A'_{11}(t)\langle{1}|+A'_{12}(t)\langle{2}|+\cdots+A'_{1n}(t)\langle{n}|,\cr
\langle\widehat{{\phi_{2}}}(t)|&=&A'_{21}(t)\langle{1}|+A'_{22}(t)\langle{2}|+\cdots+A'_{2n}(t)\langle{n}|,\cr
&&\vdots\cr
\langle\widehat{{\phi_{n}}}(t)|&=&A'_{n1}(t)\langle{1}|+A'_{n2}(t)\langle{2}|+\cdots+A'_{nn}(t)\langle{n}|,
\end{eqnarray}
where $\langle{l}|$ ($l$=$1,2,3\cdots n$) also is the bare state
for the system and $A'_{jk}(t)$ ($j,k$=$1,2,3\cdots n$) is a
devisable function associated with the state $\langle{k}|$ in
$\langle\widehat{{\phi_{j}}}(t)|$. Let's introduce two matrix
constructed by $A_{jk}(t)$ and $A'_{jk}(t)$, respectively,
\begin{eqnarray}\label{eq2-3}
A^{T}(t)= \left(\begin{array}{ccccccc}
A_{11}(t) & A_{21}(t) & \cdots & A_{n1}(t)\\
A_{12}(t) & A_{22}(t) & \cdots & A_{n2}(t)\\
\vdots & \vdots & \cdots & \vdots\\
A_{1n}(t) & A_{2n}(t) & \cdots & A_{nn}(t)\\
\end{array}\right), \cr
A'(t)= \left(\begin{array}{ccccccc}
A'_{11}(t) & A'_{12}(t) &\cdots  & A'_{1n}(t)\\
A'_{21}(t) & A'_{22}(t) & \cdots & A'_{2n}(t)\\
\vdots & \vdots & \cdots & \vdots\\
A'_{n1}(t) & A'_{n2}(t) & \cdots & A'_{nn}(t)\\
\end{array}\right),
\end{eqnarray}
where the superscript $T$ denotes the transpose operator.

In order to satisfy the biorthogonality relation and the closure
relation as shown in Eq.~(\ref{eq0-3}) and Eq.~(\ref{eq0-4}),
$A(t)$ and $A'(t)$ should satisfy following relation
\begin{eqnarray}\label{eq2-4}
  A'(t)\cdot A(t)=A^{T}(t)\cdot A'^{T}(t)=(A'(t)\cdot A(t))^{T}=\mathbf{1}_{n},
\end{eqnarray}
where $\mathbf{1}_{n}$ is the $n$-dimension unit matrix. We can
easily verify that Eq.~(\ref{eq2-4}) will be satisfied if $ A(t)$
is the reverse matrix of $A{'}(t)$.  That is, we just need to make
sure $A'(t)$ or $A(t)$ is the invertible matrix. Therefore,
mathematically,  the determinant of $A'(t)$ and $A(t)$ should
never be zero for the reverse engineered biorthogonal partners.

\subsection{Engineering quantum states by the reverse engineering method}

As an example, we now demonstrate how to engineer quantum state of
a single qubit by means of the reverse engineering method. For the
sake of simplicity,  we assume the eigenstates
$\{|{\phi_{n}}(t)\rangle\}$ of $H(t)$ read
\begin{eqnarray}\label{eq2-5}
|\phi_{1}(t)\rangle&=&A_{11}(t)|1\rangle+A_{21}(t)|2\rangle,\cr
|\phi_{2}(t)\rangle&=&A_{12}(t)|1\rangle+A_{22}(t)|2\rangle.
\end{eqnarray}
The choice of coefficients $A_{11}(t)$ and $A_{21}(t)$ is various,
we can choose the interested state as the target state
$|\phi_{1}(t)\rangle$. Without loss of generality,  by setting
$A_{11}(t)$=$-\lambda(t)\sin{\alpha(t)}$,
$A_{21}(t)$=$\cos{\alpha(t)}$,
$A_{12}(t)$=$\lambda(t)\cos{\alpha(t)}$, and
$A_{22}(t)$=$\sin{\alpha(t)}$, we can obtain
\begin{eqnarray}\label{eq2-6}
A(t)=\left(\begin{array}{ccccccc}
-\lambda(t)\sin{\alpha(t)} & \lambda(t)\cos{\alpha(t)} \\
\cos{\alpha(t)} & \sin{\alpha(t)} \\
\end{array}\right),
\end{eqnarray}
where $\lambda(t)$ and $\alpha(t)$ are time-dependent complex
functions. Obviously, $A(t)$ will be an invertible matrix if
$\lambda(t)$$\neq$$0$ is established all the time. Then, we can
obtain the accurate solution of $A^{'}(t)$
\begin{eqnarray}\label{eq2-7}
A{'}(t)= \left(\begin{array}{ccccccc}
\frac{-1}{\lambda(t)}\sin{\alpha(t)} & \cos{\alpha(t)} \\
\frac{1}{\lambda(t)}\cos{\alpha(t)} & \sin{\alpha(t)} \\
\end{array}\right).
\end{eqnarray}

Now, we start to consider the general adiabatic condition for the
designed system and calculate following matrix elements,
\begin{eqnarray}\label{eq2-8}
\langle\widehat{{\phi_{1}}}(t)|\dot{\phi}_{1}(t)\rangle&=&\frac{\dot{\lambda}(t)}{\lambda(t)}\sin^{2}\alpha(t),\cr
\langle\widehat{{\phi_{2}}}(t)|\dot{\phi}_{2}(t)\rangle&=&\frac{\dot{\lambda}(t)}{\lambda(t)}\cos^{2}\alpha(t),\cr
\langle\widehat{{\phi_{2}}}(s)|\dot{\phi}_{1}(s)\rangle&=&-\dot{\alpha}-\frac{\dot{\lambda}(s)}{\lambda(s)}\sin\alpha(s)\cos\alpha(s),\cr
\langle\widehat{{\phi_{1}}}(t)|\dot{\phi}_{2}(t)\rangle&=&\dot{\alpha}-\frac{\dot{\lambda}(t)}{\lambda(t)}\sin\alpha(t)\cos\alpha(t).
\end{eqnarray}
Then, $|{\phi_{1}}(t)\rangle$ will adiabatically evolve if the
propagator $g(t,s)$=$0$ or $g(t,s)$ is factored by a rapid
oscillating function. Mathematically, the simplest choice is
setting
$\langle\widehat{{\phi_{1}}}(t)|\dot{\phi}_{2}(t)\rangle$=$0$ (we
also can set
$\langle\widehat{{\phi_{2}}}(s)|\dot{\phi}_{1}(s)\rangle$=$0$),
and $\lambda(t)$ can be solved as
\begin{eqnarray}\label{eq2-9}
\lambda(t)=\tan\alpha(t),
\end{eqnarray}
where $\alpha(t)$$\neq$$\eta\pi/2$, $\eta$$\in$$Z$. Here, we
should note that $g(t,s)$ will also be factored by a rapid
oscillating function if $\lambda(t)$ is a constant and
$\dot{\alpha}$$\approx$$0$. In fact, this kind of setting was
examined in detail in Ref.~\cite{Mugapra2014} by Ib\'{a}\~{n}ez
and Muga. However, the weakness of this kind of setting is quite
obvious, the target state $|\phi_{1}(t)\rangle$ could not be
engineered to reach an arbitrary target state in a short time as
$\dot{\alpha}$$\approx$$0$. For the sake of generality and giving
more choices for the realization of QSE, $\lambda(t)$ will be
chosen as Eq.~(\ref{eq2-9}) in the paper. Up till now, we have
successfully completed the eigenvectors designs and obtained
following unnormalized eigenvectors
\begin{eqnarray}\label{eq2-10}
|\phi_{1}(t)\rangle&=&-\frac{\sin^{2}{\alpha(t)}}{\cos{\alpha(t)}}|1\rangle+\cos{\alpha(t)}|2\rangle,\cr
|\phi_{2}(t)\rangle&=&\sin{\alpha(t)}|1\rangle+\sin{\alpha(t)}|2\rangle.
\end{eqnarray}
According to Eq.~(\ref{eq0-5}), the system Hamiltonian   takes the
form
\begin{eqnarray}\label{eq2-11}
H(t)= \left(\begin{array}{ccc}
E_{1}(t)+\Delta_{E}(t)\cos^{2}\alpha(t)  & \Delta_{E}(t)\sin^{2}\alpha(t) \\
\Delta_{E}(t)\cos^{2}\alpha(t) & E_{1}(t)+\Delta_{E}(t)\sin^{2}\alpha(t) \\
\end{array}\right),
\end{eqnarray}
where $\Delta_{E}(t)$$\equiv $$E_{2}(t)-E_{1}(t)$ is eigenvalue
difference of the system, and it can not  equal  zero due to the
nondegeneracy.

Until now, the eigenvalues of NH Hamiltonians  are still
undetermined, although the eigenvectors designs are completed.  We
should reconsider and modify the eigenvalues resorting to
normalization ambiguities in the eigenvectors of NH Hamiltonians.
One can find that following states are also the eigenvectors of
Eq.~(\ref{eq2-11}) with the same eigenvalues,
\begin{eqnarray}\label{eq2-12}
|\phi'_{1}(t)\rangle&=&f_{1}(t)|{\phi_{1}}(t)\rangle,\cr
|\phi'_{2}(t)\rangle&=&f_{2}(t)|{\phi_{2}}(t)\rangle,
\end{eqnarray}
where $f_{1}(t)$  and $f_{2}(t)$ can be arbitrary non-zero
functions. Then, the biorthogonal partners of
$\{|{\phi'_{1}}(t)\rangle, |{\phi'_{2}}(t)\rangle\}$ read
\begin{eqnarray}\label{eq2-13}
\langle\widehat{{\phi'_{1}}}(t)|&=&\frac{1}{f_{1}^{*}(t)}\langle\widehat{{\phi_{1}}}(t)|,\cr
\langle\widehat{{\phi'_{2}}}(t)|&=&\frac{1}{f_{2}^{*}(t)}\langle\widehat{{\phi_{2}}}(t)|.
\end{eqnarray}
By calculating, we can find  the propagator $g(t,s)$ is also
factored by a rapid oscillating function for the new eigenvector
$|\phi'_{1}(t)\rangle$. That is, $|\phi'_{1}(t)\rangle$ will
continue to evolve adiabatically in current system without
additional Hamiltonians,  even though $f_{1}(t)$ is an arbitrary
non-zero function. Substituting $|\phi'_{1}(t)\rangle$ into
Eq.~(\ref{eq0-8}), we obtain
\begin{eqnarray}\label{eqx-1}
|\phi(t)\rangle=e^{i\beta'_{{1}}(t)}|{\phi'_{1}}(t)\rangle
=e^{i\beta_{{1}}(t)}f_{1}(0)|{\phi_{1}}(t)\rangle,
\end{eqnarray}
where the adiabatic phase for the new eigenvector reads
\begin{eqnarray}\label{eqx-2}
\beta'_{1}(t)&=&\int_{0}^{t}[-E_{1}(t')+i\langle\widehat{{\phi'_{1}}}(t')|{\dot{\phi}'_{1}}(t')\rangle]dt'\cr
&=&\int_{0}^{t}[-E_{1}(t')+i\langle\widehat{{\phi_{1}}}(t')|{\dot{\phi}_{1}}(t')\rangle
+i\textrm{d}\ln{f_{1}(t)}]dt'.
\end{eqnarray}
As a consequence, the normalization ambiguities in the
eigenvectors only generates a constant multiplication factor
$f_{1}(0)$,  and the target state $|\phi_{1}(t)\rangle$ always
 evolves adiabatically in current system. Furthermore, when
the auxiliary adiabaticity condition with respect to  the
adiabatic phase [see Eq.~(\ref{eq0-10})] is taken into account,
\begin{eqnarray}\label{eqx-3}
\textrm{Im}[E_{1}(t)]
=\textrm{Re}[\langle\widehat{{\phi_{1}}}(t)|{\dot{\phi}}_{1}(t)\rangle]
=\textrm{Re}[\sin\alpha(t)\cos\alpha(t)],
\end{eqnarray}
the target state won't suffer strong exponential variations which
is remarkable for quantum information processing.

We can find that Eq.~(\ref{eq2-11}) can be expressed in terms of
the Pauli matrices as
\begin{eqnarray}\label{eqx-4}
H(t)=\frac{\Delta_{E}(t)}{2}\sigma_{x}-i\frac{\delta(t)}{2}
\sigma_{y}+\frac{\delta(t)}{2}\sigma_{z}+E'_{0}(t)\mathbf{1},
\end{eqnarray}
where $\delta(t)$=$\Delta_{E}(t)\cos(2\alpha(t))$ and
$E'_{0}(t)$=$E_{1}(t)+{\Delta_{E}(t)}/{2}$ are the time-dependent
variables, and $\mathbf{1}$ denotes the unit matrix. In fact, the
real part of $E'_{0}(t)$ can be ignored by applying appropriate
energy shift, which doesn't play a negative role in the
investigation of population of system. The system can be mapped
onto the Hamiltonian
\begin{eqnarray}\label{eqx-5}
H(t)=\frac{1}{2}[\Delta_{E}(t)\sigma_{x}-i\delta(t)
\sigma_{y}+\delta(t)\sigma_{z}]+i~\textrm{Im}[E'_{0}(t)]\mathbf{1}.
\end{eqnarray}
It can be easily found there are only two variables,
$\Delta_{E}(t)$ and $\alpha(t)$, in Eq.~(\ref{eqx-5}). Thus, the
crucial NH Hamiltonian engineering can be cast into the
$\Delta_{E}(t)$ design and the ${\alpha(t)}$ design. Theoretically
speaking,  besides the consistency condition
[$\alpha(t)$$\neq$$2\eta/\pi$, $\Delta_{E}(t)$$\neq$$0$ and
$\sin\alpha(0)$$\approx$$1$ (it should be noted that the initial
state could make  connection with the target state
$|\phi_1(t)\rangle$ by setting $\sin\alpha(0)$$\approx$$1$
according to Eq.~(\ref{eq2-10}))], there is almost no limit on the
choices of $\Delta_{E}(t)$ and ${\alpha(t)}$ for engineering the
system to reach an arbitrary target state at a predefined time.
However, the choices of $\alpha(t)$ and $\Delta_{E}(t)$ will
affect  evolution speed for the target state and the feasibility
in the practical realization. Especially, when the term
$\textrm{Im}[E'_{0}(t)]$ in Eq.~(\ref{eqx-5}) does not equal zero,
the practical realization of this Hamiltonian is significantly
challenged in  experiments. We shall explore in the following
subsection to find an appropriate physical model that can
incorporate the resulting Hamiltonian.

\section{Experimental Feasibility and Numerical Examples}\label{section:IV}
For the purpose of convenience,  we consider a simple case of
Eq.~(\ref{eqx-5}),
\begin{eqnarray}\label{eq3-1}
\textrm{Im}[E'_{0}(t)]=\textrm{Im}[E_{1}(t)]+\textrm{Im}[\frac{\Delta_{E}(t)}{2}]\approx0.
\end{eqnarray}
The Hamiltonian of system  reduces
\begin{eqnarray}\label{eq3-2}
H(t)=\frac{1}{2}[\Delta_{E}(t)\sigma_{x}-i\delta(t)
\sigma_{y}+\delta(t)\sigma_{z}].
\end{eqnarray}

In general, there is no simple ``real'' field interaction leading
to Eq.~(\ref{eq3-2}), since the off-diagonal terms of the
resulting Hamiltonian are different. For example, we assume a
semiclassical description of the interaction between a ``real''
magnetic field $\textit{\textbf{B}}(t)$ and a rotating spin qubit,
where $\textit{\textbf{B}}(t)$=$[B_{x}(t)e_{x}$+$B_{y}(t)
e_{y}$+$B_{z}(t)e_{z}]/2M_{b}$,  $e_{r}$ $(r$=$x,y,z)$ is the unit
vector along $r$ axie, $M_{b}$=$\hbar e/(2m)$ is the Bohr
magneton, and $B_{r}(t)$ is real variable. Then, the Hamiltonian
of this system reads
\begin{eqnarray}\label{eq1}
H(t)= \left(\begin{array}{ccc}
B_{z}(t)  & B_{x}-iB_{y}(t) \\
B_{x}(t)+iB_{y}(t) & -B_{z}(t) \\
\end{array}\right),
\end{eqnarray}
we can find that the off-diagonal terms  are complex conjugate of
each other which does meet the requirements. However, we may
obtain the resulting Hamiltonian if the magnetic field
$\textit{\textbf{B}}(t)$ is the complex signal field rather than
the real signal field, for example,
\begin{eqnarray}\label{eq2}
B_{x}(t)\rightarrow A_x(t)e^{I\Theta_x(t)}&=&\textrm{Re}[\Delta
E(t)]+i \textrm{Im}[\Delta E(t)], \cr B_{y}(t)\rightarrow
A_y(t)e^{I\Theta_y(t)}&=&\textrm{Im}[\delta(t)]-i
\textrm{Re}[\delta(t)], \cr B_{z}(t)\rightarrow
A_z(t)e^{I\Theta_z(t)}&=&\textrm{Re}[\delta(t)]+i
\textrm{Im}[\delta(t)],
\end{eqnarray}
where $A_{r}$ is the amplitude and $\Theta_{r}$ is the phase. In
fact, a similar complex signal field has been discussed in detail
in Refs.~\cite{Cohen1995,Muga2011} and references therein.
Additionally, the phase $\Theta_{r}$ can also be considered as the
dissipation factor which is introduced by the noise (e.g. the
dephasing effects  due to the collisions or phase fluctuations of
the magnetic fields or when the rotating-wave approximation fails
for the strong magnetic fields~\cite{Muga2011}). Therefore, the
resulting Hamiltonian Eq.~(\ref{eq3-2}) is accessible
experimentally with the complex signal field or the real signal
field under some dissipation effects.

Now, let's focus on how to design $\Delta_{E}(t)$ and
${\alpha(t)}$  from an experimental view point. At first, we can
write ${\alpha(t)}$ in polar form
\begin{eqnarray}\label{eq3-3}
{\alpha(t)}=\rho(t)\exp(i\theta(t)),
\end{eqnarray}
where $\rho(t)$ and $\theta(t)$ are time-dependent real variables.
It is useful to rewrite $\Delta_{E}(t)$, taking into account
Eq.~(\ref{eqx-3}) and Eq.~(\ref{eq3-1}), as
\begin{eqnarray}\label{eq3-4}
\Delta_{E}(t)=\textrm{Re}[\Delta_{E}(t)]-i\sin[2\rho(t)
\cos\theta(t)]\cosh[2\rho(t) \sin\theta(t)],
\end{eqnarray}
where the real part of $\Delta_{E}(t)$ is a undertermined
parameter and  the selection of $\textrm{Re}[\Delta_{E}(t)]$ seems
quite arbitrary mathematically.  However, $\Delta_{E}(t)$ is
physically associated with the eigenvalue difference of the system
[see Eq.~(\ref{eq2-11})]. Thus, we should guarantee the modulus of
$\textrm{Re}[\Delta_{E}(t)]$ is relatively large, otherwise, the
system will undergo transitions between $|\phi_{1}(t)\rangle$ and
$|\phi_{2}(t)\rangle$ constantly. Furthermore, $\Delta_{E}(t)$ is
also associated with the magnetic field, we should consider the
experimental technology for the magnetic field engineering. Once
$\textrm{Re}[\Delta_{E}(t)]$, $\rho(t)$, and $\theta(t)$ are
fixed, which means the magnetic fields $\textbf{\textit{B}}$ is
fixed. However, it should be emphasized that an arbitrary choice
of $\rho(t)$ and $\theta(t)$ will typically lead to singularities
on the magnetic field. We will detailedly discuss this problem in
following physical model.

In the above derivation, we have considered a simple case of
Eq.~(\ref{eqx-5}), that is, $\textrm{Im}[E'_{0}(t)]$$\approx$$0$.
Now, we will discuss the experimental feasibility for the physical
model when $\textrm{Im}[E'_{0}(t)]$$\gg$$0$. For convenient
discussion, we assume $\textrm{Im}[\delta(t)]$=$\Gamma(t)$, where
$\Gamma(t)$ is a time-dependent real coefficient. In this case,
the Eq.~(\ref{eqx-5}) can be written as
\begin{eqnarray}\label{eq3-5}
H(t)&=&\frac{1}{2}[\Delta_{E}(t)\sigma_{x}-i\delta(t)
\sigma_{y}+\textrm{Re}[\delta(t)]\sigma_{z}]\cr
&&+i\left(\begin{array}{ccc}
\textrm{Im}[E'_{0}(t)]+\frac{ \Gamma(t)}{2}  & 0 \\
0 & \textrm{Im}[E'_{0}(t)]-\frac{ \Gamma(t)}{2} \\
\end{array}\right).
\end{eqnarray}
Note that the difference of the order of magnitude between
$\textrm{Im}[E'_{0}(t)]$ and $\Gamma(t)$ is little, otherwise, the
problem seems to be equivalent to  above simple example. More
specially, setting
\begin{eqnarray}\label{eq3-6}
\textrm{Im}[E'_{0}(t)]+ \Gamma(t)/{2}=0,
\end{eqnarray}
 we will find the resulting Hamiltonian in Eq.~(\ref{eq3-5})
can be accessible  in the following physically setting: a spin
qubit or atom passes through a region of rapidly varying magnetic
field $\textit{\textbf{B}}$=$[{\Delta_{E}(t)}e_{x}-i\delta(t)
e_{y}$+$\textrm{Re}(\delta(t))e_{z}]/2M_{b}$, and the spin qubit
or atom suffers a radiation process with the dissipation rate
\cite{Torosov2013,ptduichen,Mugajpb2008} $\Gamma(t)$ (e.g. the
spontaneous decay, in some cases, $\Gamma(t)$ can be controlled as
an effective decay rate by further interactions, see, e.g.,
Ref.~\cite{Mugajpb2008}). This is remarkable,  since the noise and
certain dissipation in the systems are no longer undesirable, but
play an integral part in our scheme.

From an experimental view point, we should consider the
$\Delta_{E}(t)$ design and the ${\alpha(t)}$ design for current
physical model.  Similar to above derivation, ${\alpha(t)}$ is
still in polar form. Substituting Eqs.~(\ref{eqx-3}) and
(\ref{eq3-3}) into Eq.~(\ref{eq3-6}), we will find $\Delta_{E}(t)$
satisfies following equation
\begin{eqnarray}\label{eq3-7}
-\textrm{Re}[\Delta_{E}(t)]\Omega_{1}=\textrm{Im}[\Delta_{E}(t)](1+\Omega_{2})+\Omega_{3},
\end{eqnarray}
where
\begin{eqnarray}\label{eq3-8}
\Omega_{1}&=&\sin[2\rho(t)\cos\theta(t)]\sinh[-2\rho(t)\sin\theta(t)],\cr
\Omega_{2}&=&\cos[2\rho(t)\cos\theta(t)]\cosh[2\rho(t)\sin\theta(t)],\cr
\Omega_{3}&=&\sin[2\rho(t)\cos\theta(t)]\cosh[2\rho(t)\sin\theta(t)].
\end{eqnarray}
Furthermore, $\Gamma(t)$ can be simplified as
$\Gamma(t)$=$-\textrm{Im}[\Delta_{E}(t)]-\Omega_{3}$. Apparently,
once $\textrm{Im}[\Delta_{E}(t)]$ is specified, the magnetic field
$\textbf{\textit{B}}$ and $\Gamma(t)$ are straightforwardly
calculated with Eqs.~(\ref{eq3-7}) and (\ref{eq3-8}). On the other
hand, the form of $\textrm{Im}[\Delta_{E}(t)]$ can be derived with
the inversion strategy, if the form of dissipation rate
$\Gamma(t)$ is fixed. This is remarkable, since we can choose
appropriate extra magnetic fields to adiabatically drive an
artificial quantum state for certain dissipative quantum system.
Up to now, we have in principle constructed the magnetic fields
according to the ${\alpha(t)}$ design and specified dissipation
rate $\Gamma(t)$. However, the ${\alpha(t)}$ designs are
problematic, as an arbitrary choice of $\rho(t)$ and $\theta(t)$
will typically lead to singularities on the right-hand side of
Eq.~(\ref{eq3-8}) (for instance, $\Omega_{i}$ ($i$=$1,2,3$) will
jump abruptly when $2\rho(t)\cos\theta(t)$=$\eta\pi$ or
$2\rho(t)\sin\theta(t)$$\approx$$0$). In general, $\Omega_{i}$
will also introduce singularities in magnetic fields, then, we
could not construct the finite and smooth magnetic fields. Thus,
we should design $\rho(t)$ and $\theta(t)$ to avoid the
singularities. It is advisable to fix $\rho(t)$ or $\theta(t)$
first, then design the other one to avoid the singularities. A
simple example is
\begin{eqnarray}\label{eq3-9}
\rho(t)=\frac{\pi}{2}-o-\xi\sin\mu t,\ \theta(t)=\zeta+\sin\nu t,
\end{eqnarray}
where $\mu$ and $\nu$ are constant frequencies related to the
concrete phase engineering, and $o$ is an extremely small constant
to keep the consistency condition. By choosing appropriate
parameters (such as $\xi$=$0.4\pi $ $\zeta$=$0.08\pi$, and
$\mu$=$\nu$=$0.5\Omega$), we can construct the finite and smooth
magnetic fields.
\begin{figure}
\scalebox{0.3}{\includegraphics{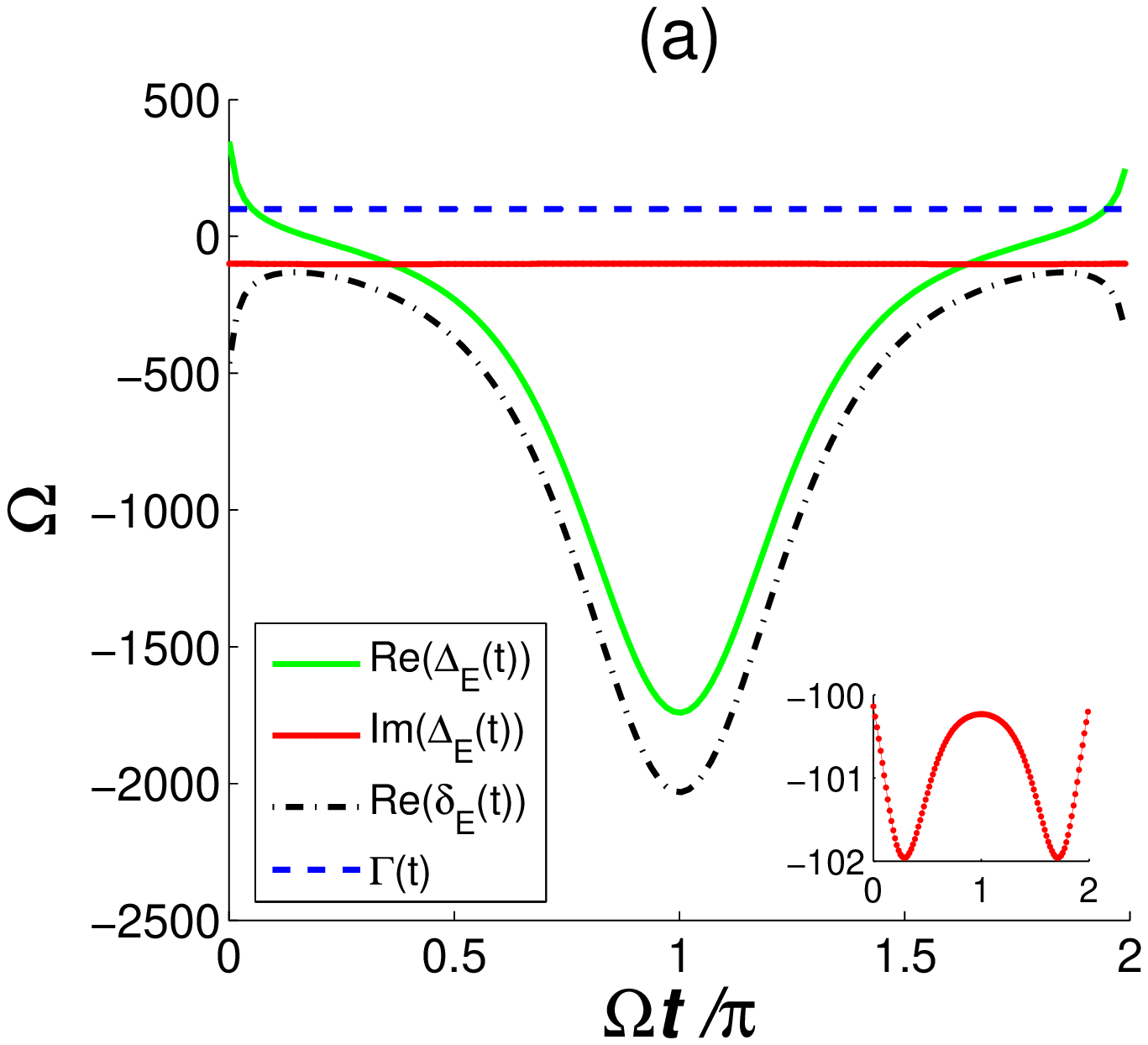}\includegraphics{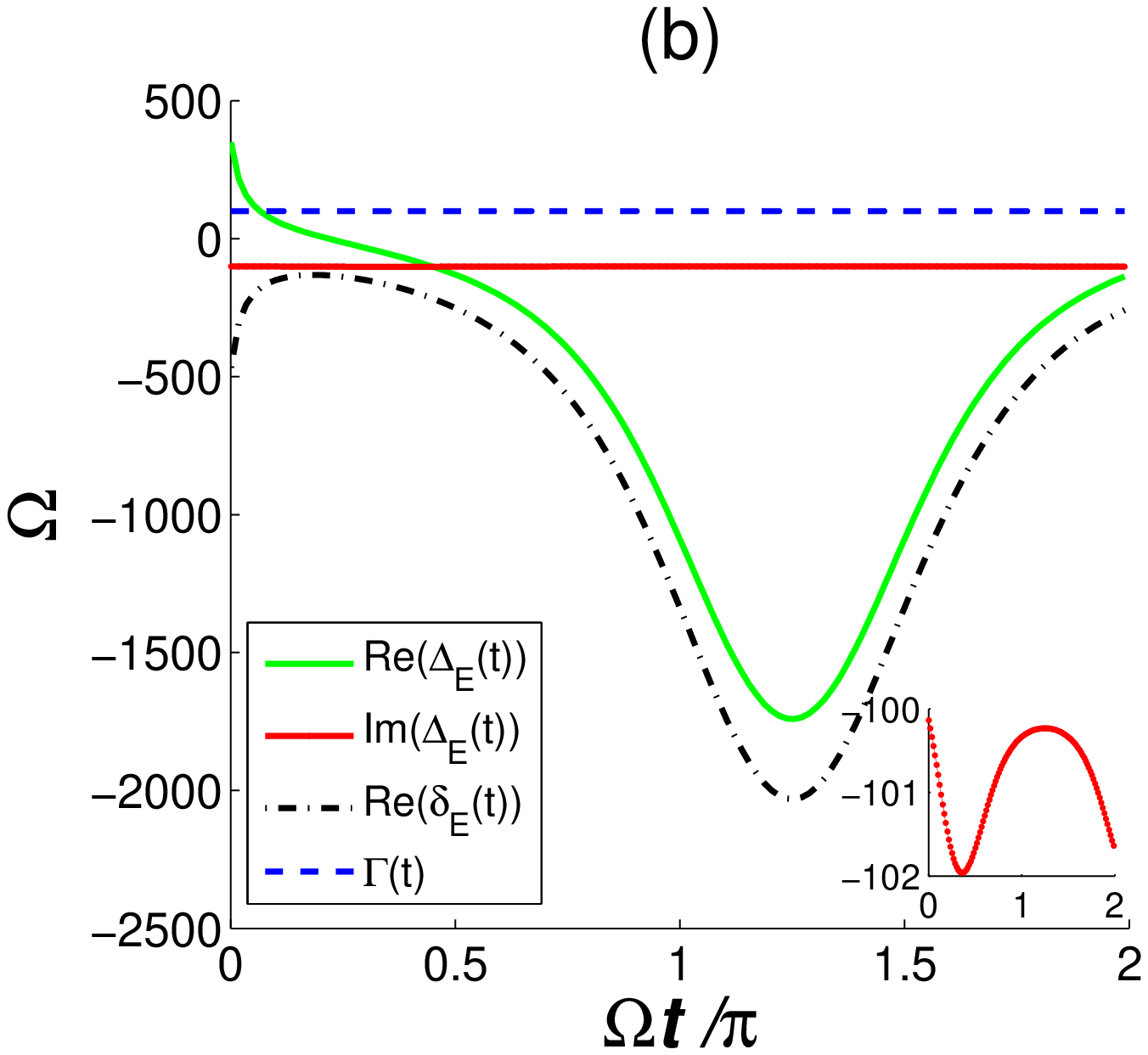}}
\scalebox{0.3}{\includegraphics{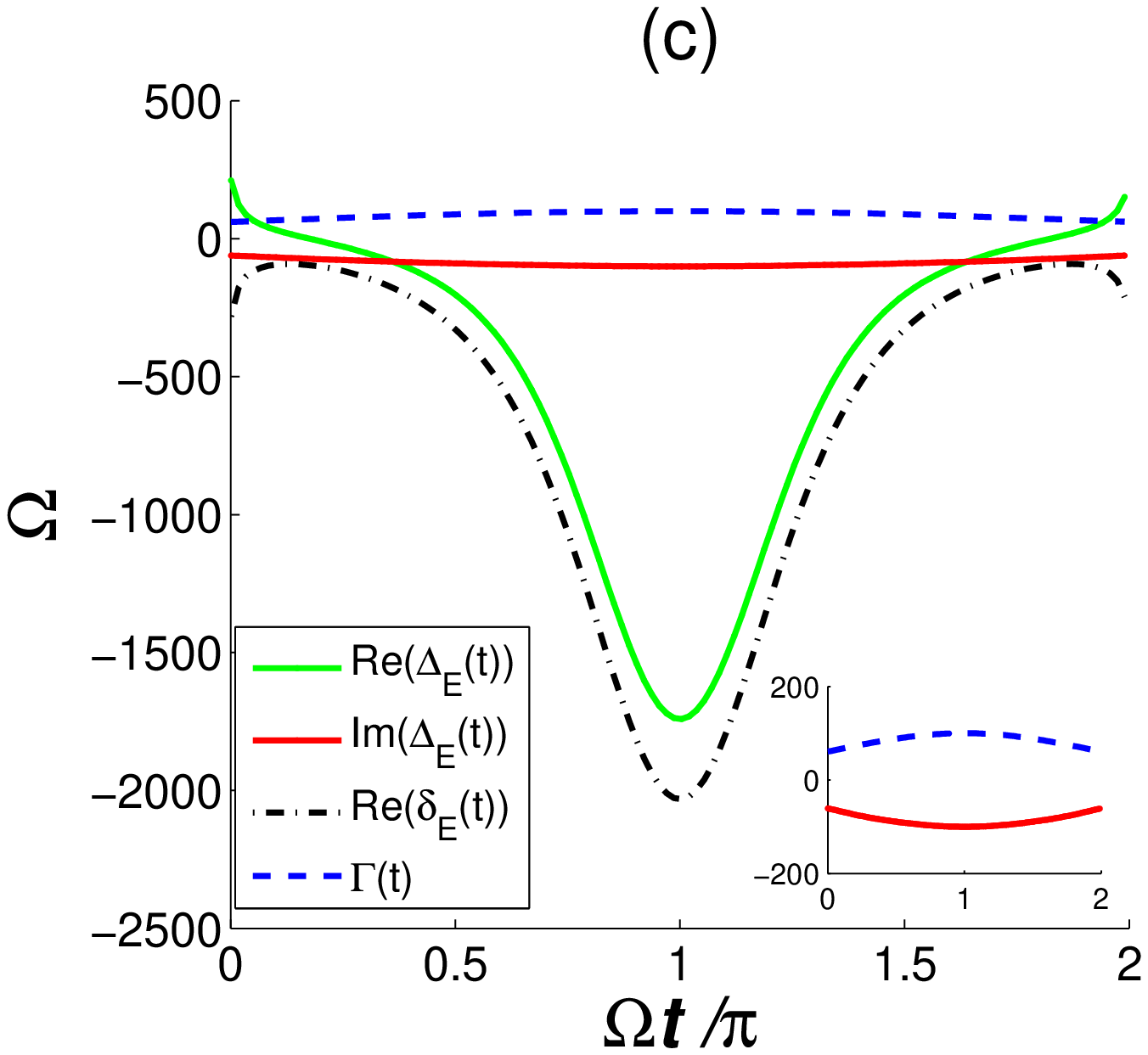}\includegraphics{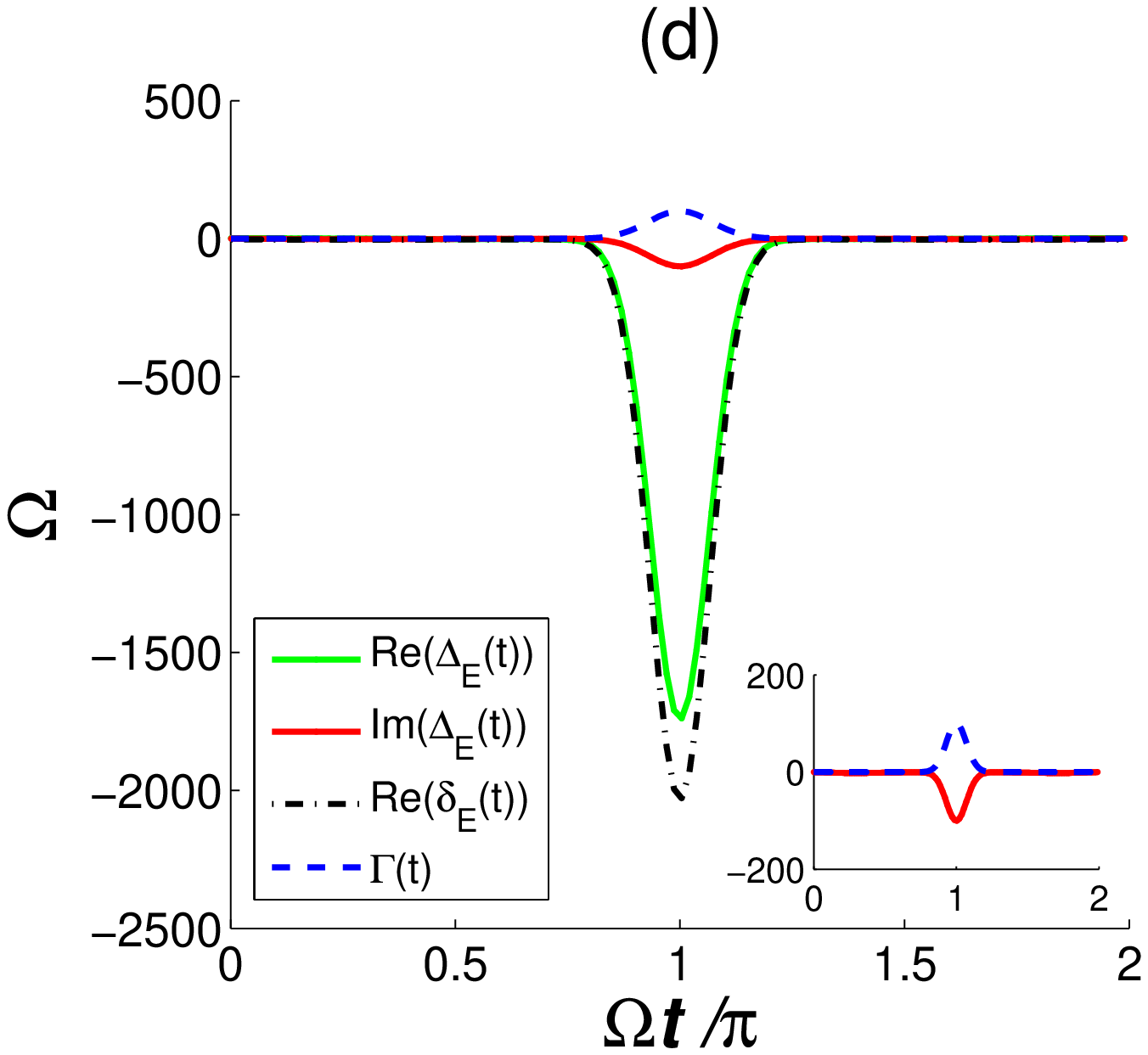}}
\caption{\label{fig1} Time evolution of the magnetic fields with
different parameters. For (a) and (b) the dissipation rate
$\Gamma(t)$=$100\Omega$: (a) $\mu$=$\nu$=$0.5\Omega$, (b)
$\mu$=$\nu$=$0.4\Omega$ (the other parameters are $\xi$=$0.4\pi$
and $\zeta$=$0.08\pi$). For (c) and (d) the dissipation rate
$\Gamma(t)$ is based on Eq.~(\ref{eq3-10}): (c)
$T$=$\sqrt{2/\Omega}$, (d) $T$=$\sqrt{0.01/\Omega}$ (the other
parameters are $\Omega'$=$100\Omega$, $t_{0}$=$\pi/\Omega$,
$\mu$=$\nu$=$0.5\Omega$, $\xi$=$0.4\pi$, and $\zeta$=$0.08\pi)$. }
\end{figure}
For an intuitive grasp of the change of magnetic fields with
different parameters in the dissipation system, we plot the time
evolution of magnetic fields in Fig.~\ref{fig1}. As shown in
Fig.~\ref{fig1}(a), when the dissipation rate $\Gamma(t)$ is a
constant ($\Gamma(t)$=$100\Omega$), the shape of the magnetic
fields are not very complex and the maximum value of the magnetic
fields ($\Omega_\textrm{{max}}$) is about $2000\Omega$. From an
experimental view point, if $\Omega$=$2\pi\times 10$ KHz,
$\Omega_\textrm{{max}}$ is about $2\pi\times 20$ MHz, which is
feasible with present experimental
techniques~\cite{Bloch1946,Sigmund2001,Goodson2002,Kong2014,Jiang2007,Vandersypen2001}.
Thus, the magnetic fields in our scheme are not hard to be
realized in practice. Figs.~\ref{fig1}(a) and \ref{fig1}(b) share
the same dissipation rate $\Gamma(t)$, while the parameters in
$\rho(t)$ and $\theta(t)$ are different.  In fact, we are also
interested in the time evolution of magnetic fields for a
dissipation system with a time-dependent dissipation rate
$\Gamma(t)$. Without loss of generality, we take a Gaussian
dissipation rate $\Gamma(t)$ as an example
\begin{eqnarray}\label{eq3-10}
\Gamma(t)=\Omega'e^{[-(\frac{t-t_{0}}{T})^{2}]},
\end{eqnarray}
where $\Omega'$ is a constant frequency, while $T$ and $t_{0}$ are
time constants. We should emphasize that $T$ is related to the
time scale of $\Gamma(t)$ physically, it should be chosen
appropriately to keep the validity of the noise or certain
dissipation. Figures~\ref{fig1}(c) and \ref{fig1}(d) display the
time evolution of magnetic fields with different $T$, while the
others parameters are identical. Apparently, the magnetic fields
in Figs.~\ref{fig1}(b) and \ref{fig1}(c) are similar to the
magnetic fields in Fig.~\ref{fig1}(a), all of them are feasible in
practice. However, the magnetic fields in Fig.~\ref{fig1}(d) are
quite different from others. We can clearly see that magnetic
fields in most of time can be neglected, specially,
$\Delta_{E}(t)$ is equal to zero for a long time, which means the
invalidity of the consistency condition ($\Delta_{E}(t)\neq0$,
since $\Delta_{E}(t)$ is also associated with the difference
between the eigenvalues of the system physically). Therefore, the
choice of the parameters in Fig.~\ref{fig1}(d) is problematic or
false.

\begin{figure}
\scalebox{0.3}{\includegraphics{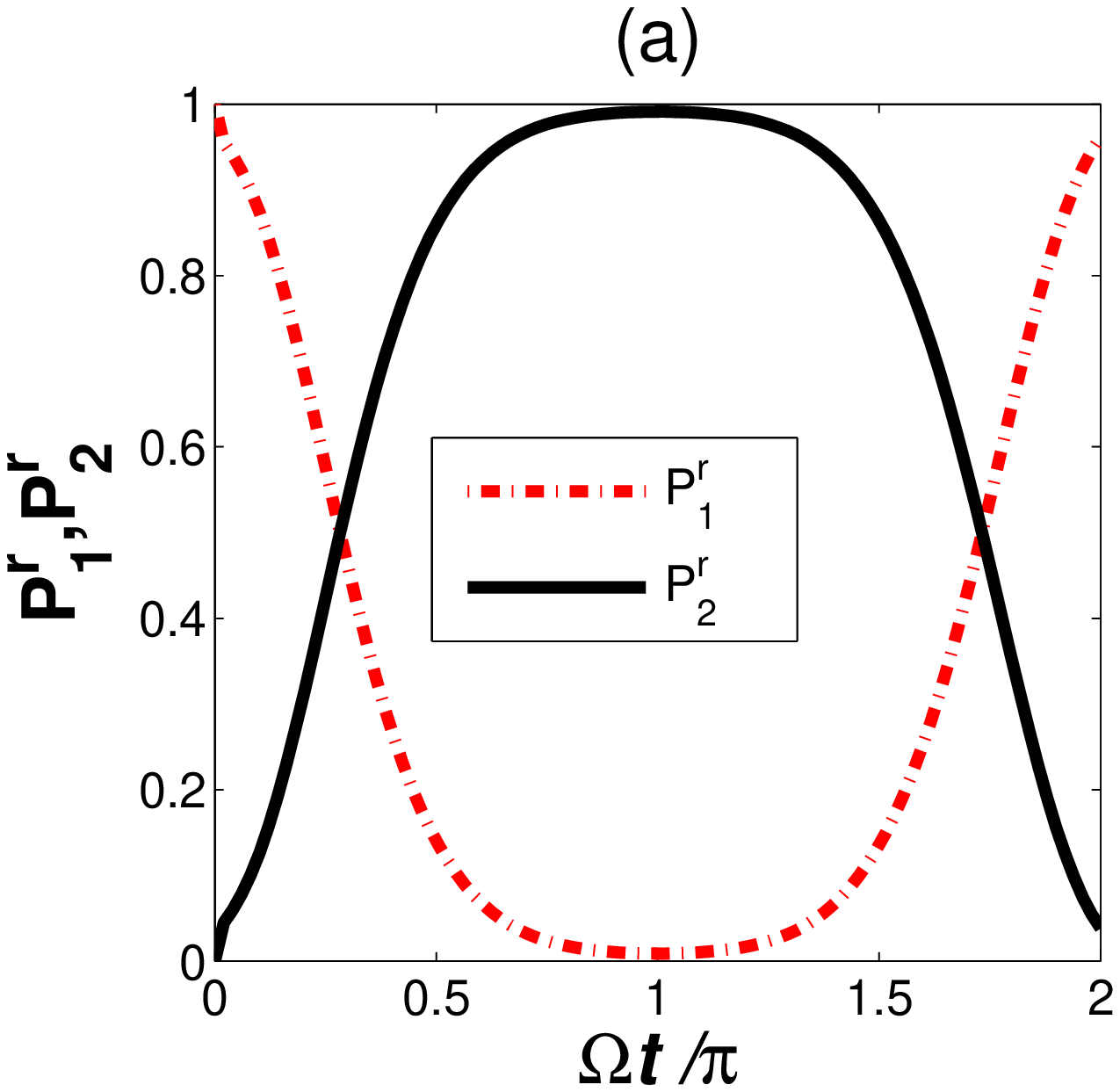}\includegraphics{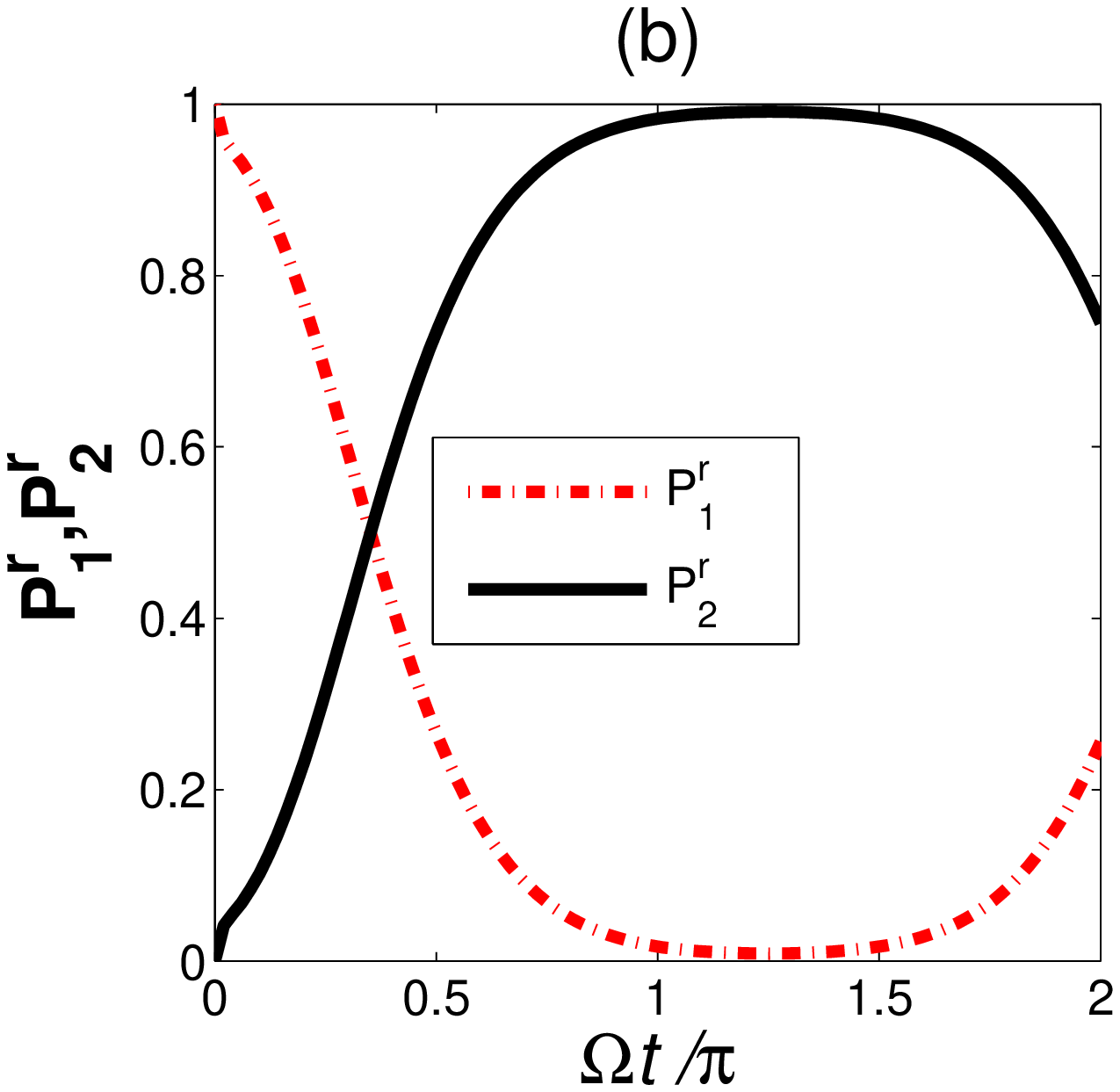}}
\scalebox{0.3}{\includegraphics{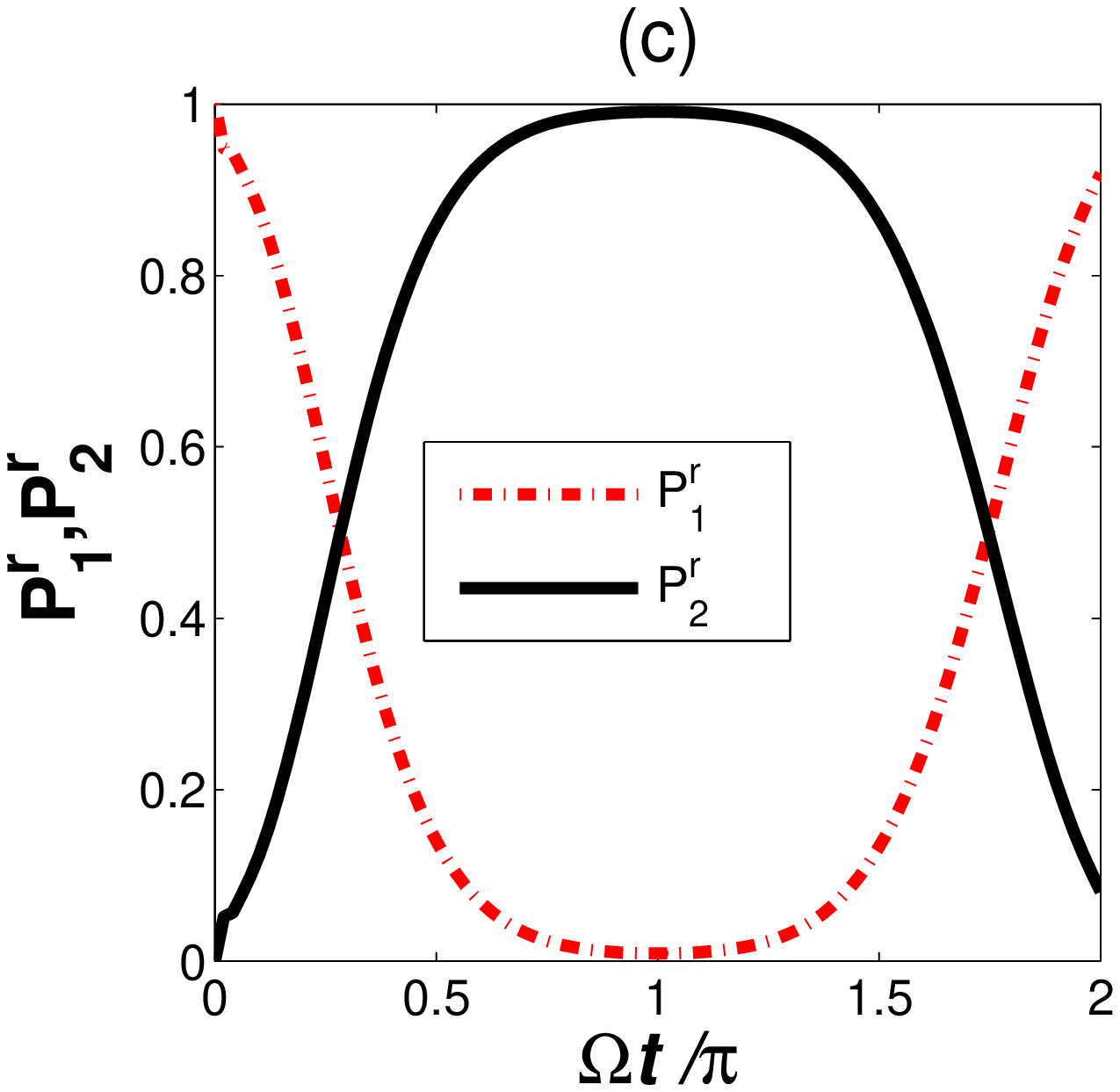}\includegraphics{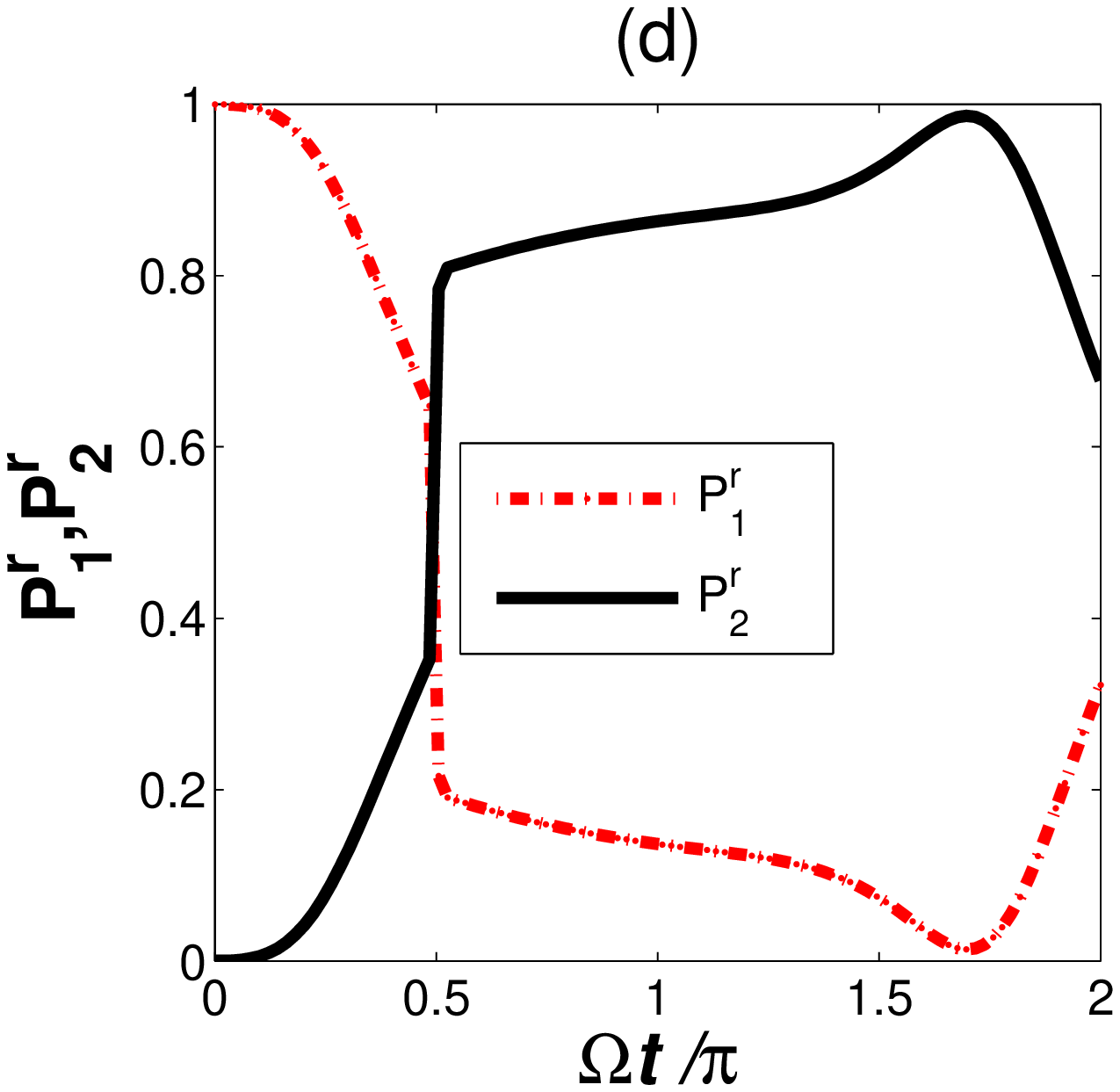}}
\caption{\label{fig2}Time evolution of the relative populations
for the states $|1\rangle$ and $|2\rangle$ with different magnetic
field parameters. The parameters are the same as shown in the
caption of Fig.~\ref{fig1}: (a) $\Gamma(t)$=$100\Omega$,
$\mu$=$\nu$=$0.5\Omega$; (b) $\Gamma(t)$=$100\Omega$,
$\mu$=$\nu$=$0.4\Omega$ ($\xi$=$0.4\pi$ and $\zeta$=$0.08\pi$);
(c) $\Gamma(t)$=$100\Omega\exp{[-({t-t_{0}})^{2}/{T}^{2}]}$,
$T$=$\sqrt{2/\Omega}$; (d)
$\Gamma(t)$=$100\Omega\exp{[-({t-t_{0}})^{2}/{T})^{2}]}$,
$T$=$\sqrt{0.01/\Omega}$ ($t_{0}$=$\pi/\Omega$,
$\mu$=$\nu$=$0.5\Omega$, $\xi$=$0.4\pi$, and $\zeta$=$0.08\pi)$.
}
\end{figure}

Now we start to study the population engineering of the bare state
in the target state. However, as shown in Eq.~(\ref{eq2-10}) and
Eq.~(\ref{eq2-12}), the target state seems to be no natural
normalization. For an intuitive grasp of the change of the
population engineering of the bare basic, we will use the relative
population ${P_{i}^{r}}$ $(i$=$1,2)$ to study the effects of
different magnetic fields on the population engineering, where the
relative population is defined as
${P_{i}^{r}}$=$P_{i}/(P_{1}$+$P_{2})$, and $P_{i}$ is the
population for the bare state $|i\rangle$. We consider a realistic
case of an extremely small population in the bare state
$|2\rangle$ for the initial state
\begin{eqnarray}\label{eq3-13}
|\phi(0)\rangle\simeq|{\phi'_{1}}(0)\rangle
=\sqrt{1-o^{2}}|1\rangle+o|2\rangle,
\end{eqnarray}
where $o$ is an extremely small constant. In Fig.~\ref{fig2}, we
plot the time evolution of the relative populations for the bare
states $|1\rangle$ and $|2\rangle$ with the same parameters as
shown in the caption of Fig.~\ref{fig1}. We can find that the
relative populations $P_{1}^{r}$ and $P_{2}^{r}$ almost have the
same evolving tendency in Figs.~\ref{fig2}(a) and \ref{fig2}(c),
and a perfect full relative population inversion is complied when
$\Omega t$=$\pi$. It should be noted that the time for a full
relative population inversion is about 50 ns which is short, if
$\Omega$=$2\pi$$\times$$ 10$ KHz. Figure~\ref{fig2}(b) also
clearly shows a full relative population inversion when $\Omega
t$$\approx $$1.3\pi$. However, the time evolution of the relative
populations in Fig.~\ref{fig2}(d) are complicated, which are quite
different from others. The reason for this result is that the
choice of the parameters in Fig.~\ref{fig2}(d) is problematic or
false, more particularly, the $T$ is too short and the consistency
condition is invalid in this case.

\begin{figure}
\scalebox{0.3}{\includegraphics{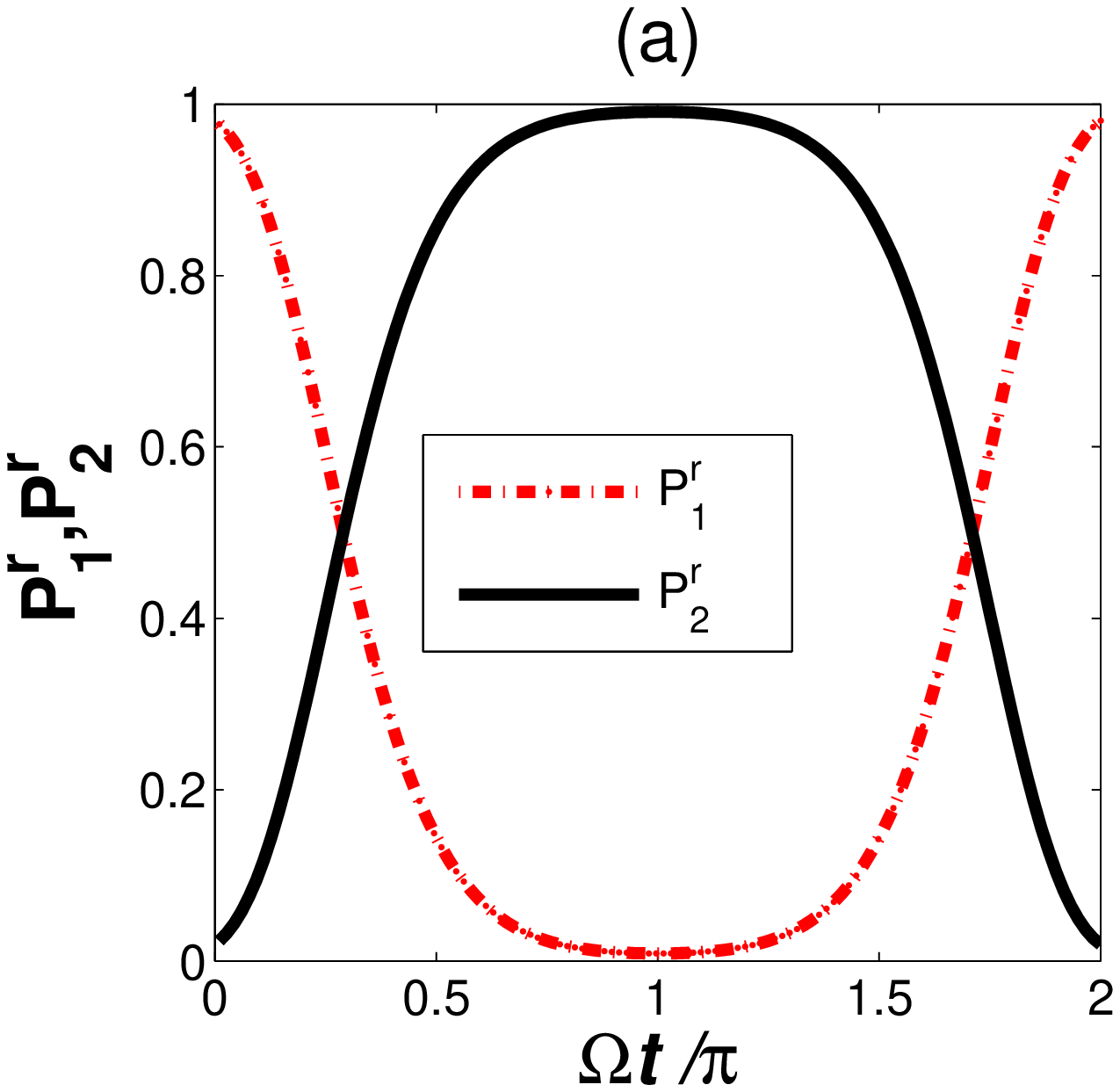}\includegraphics{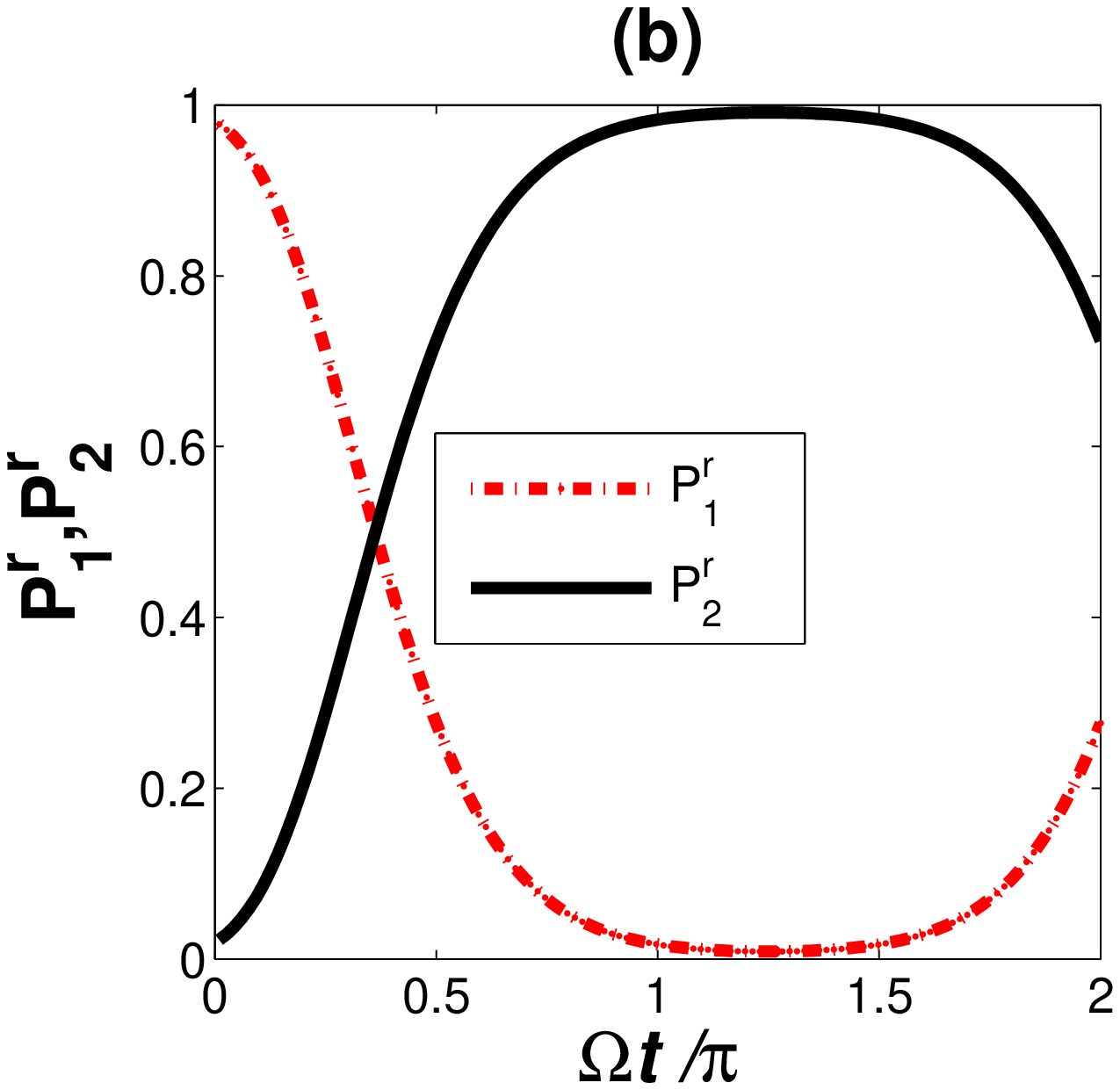}}
\caption{\label{fig3}The ideal population engineering with
different $\alpha(t)$. $\alpha(t)$ is based on Eqs.~(\ref{eq3-3})
and (\ref{eq3-9}): (a) $\mu$=$\nu$=$0.5\Omega$, $\xi$=$0.4\pi$ and
$\zeta$=$0.08\pi$; (b)
 $\mu$=$\nu$=$0.4\Omega$, $\xi$=$0.4\pi$ and
$\zeta$=$0.08\pi$. }
\end{figure}
To judge the validity of our scheme for adiabatically driving, we
should compare the real population engineering with the ideal
population engineering [see Eq.~(\ref{eq2-10})]. The ideal
population engineering with different $\alpha(t)$ are given in
Fig.~\ref{fig3}. As shown in  Eq.~(\ref{eq2-10}), the ideal
population engineering only depends on the ${\alpha(t)}$ design.
In other words, the population engineering will be identical for
the same ${\alpha(t)}$ design independent of other parameters.
Thus, if our scheme is valid, Figs.~\ref{fig2}(a), \ref{fig2}(c)
and \ref{fig3}(a) (Figs.~\ref{fig2}(b) and \ref{fig3}(b)) should
be identical, since the ${\alpha(t)}$ designs for them are
identical. Obviously, the results are consistent with our
deduction, hence our scheme can work well even under noise if the
parameters are chosen appropriately. In addition, we can get more
interested target states with different ${\alpha(t)}$ designs.

\section{DISCUSSION AND CONCLUSION}\label{section:V}

We have generalized the quantum adiabatic theorem to the NH system
and provided a strict adiabaticity condition to make the adiabatic
evolution non lossy. The strict adiabaticity condition can be
regarded as a non-trivial generalization of adiabaticity
conditions for the Hermitian Hamiltonians presented by Jing
\textit{et al.}~\cite{Jingpra2014}. According to the strict
adiabaticity condition, one should eliminate the non-adiabatic
couplings and the effect of the imaginary part of adiabatic phase
as much as possible. The NH Hamiltonian {reverse engineering}
method has been proposed to adiabatically drive an artificial
quantum state. {A concrete two-level system example was discussed
to show the usefulness of the reverse engineering method in the
paper}, and numerical simulation  showed that our scheme can work
well even under noise if the parameters are chosen appropriately.
Furthermore, we can obtain the desired target state by adjusting
extra rotating magnetic fields at a predefined time. Specifically,
the noise or certain dissipation in the systems are no longer
undesirable, but play a positive role in our scheme. Therefore,
our scheme is powerful and reliable for the quantum information
processing.

The present work bears some common elements with the quantum
control in open quantum systems, including the idea of using
dissipation as a resource (e.g. dissipative quantum dynamics
(DQD)~\cite{Verstraete2009,Vacanti2009,Kastoryano2011,shen2012}
and the NH shortcuts to adiabaticity
schemes~\cite{Muga2011,Torosov2013}), so it is worth stressing the
similarities and differences. In fact, the basic idea of DQD can
be summarized as follows: the interaction between the system and
the environment is modulated to make the target state become the
stationary state of the system. Therefore, some specific
dissipative factors are no longer undesirable, but can be regarded
as the important resources. For the NH shortcuts to adiabaticity
schemes, the dissipative factors  are also introduced to the
system to cancel somehow the non-adiabatic losses. In this way,
one can improve dramatically the fidelity of the adiabatic
passage. However, a common problem which one may encounter via DQD
or the NH shortcuts to adiabaticity is  how to  use the specific
dissipative factors or employ the appropriate interaction between
the system and the environment. Furthermore, those methods may
also be limited severely for some applications (the non-adiabatic
dynamics processes), since the starting point of them generally is
to improve a given (adiabatic) dynamics process.

Among the differences with the present
works~\cite{Dridi2012,Mugapra2014,Verstraete2009,Vacanti2009,Kastoryano2011,shen2012,Muga2011,Torosov2013},
the most prominent point is: using the reverse engineering method,
we can easily obtain the Hamiltonian to realize the intended
dynamics without loss, which allows one to design the Hamiltonian
according to the demand. The main task we should consider is how
to physically realize the resulting NH Hamiltonian. Sometimes, the
resulting NH Hamiltonian may be hard to be realized (a common
potential problem of the NH shortcuts to adiabaticity). However,
we should note that the difficulty to realize the NH Hamiltonian
may be solved by enlarging the system with the aid of Naimark
extensions~\cite{Gunther2008}. Furthermore, in a sense, all the
resulting NH Hamiltonian (even the problematic NH Hamiltonian) may
help us with a deeper understanding on the problem: which
dissipative factors are the specific dissipative factors that can
be used as a resource to realize QSE, and promote the development
of quantum information science in NH system frames.

Furthermore, any quantum system whose Hamiltonian is possible to
be simplified into the form in Eq.~(\ref{eqx-5}) (the basic for
the simplified Hamiltonian can be arbitrary dressed states, as
long as, the dressed states satisfy the biorthogonality relation
and closure relation), the scheme can be implemented
straightforward. This might lead to a useful step toward realizing
fast and noise-resistant quantum information processing for
multi-qubit systems in current technology. The applications or
extensions of this work may be in fields, such as $n$-dimensional
systems~\cite{Macchiavello2002,Cerf2002} (for instance, the
three-dimensional systems for the stimulated Raman adiabatic
passage), superadiabatic treatments \cite{Joye2007,Giannelli2014},
and non-adiabatic evolution of NH quantum systems
\cite{Torosov2013}.

\section*{ACKNOWLEDGEMENT}
This work was supported by the National Natural Science Foundation
of China under Grants No. 11575045, No. 11374054 and No. 11675046,
and the Major State Basic Research Development Program of China
under Grant No. 2012CB921601.

\end{document}